\documentclass{elsart}

\usepackage{graphicx}

\usepackage{amssymb}
\usepackage{amsmath}

\begin{document}

\begin{frontmatter}

\title{A simple model of reactor cores for reactor neutrino flux calculations for the KamLAND experiment}

\author[RCNS]{K. Nakajima\corauthref{cor1}},
\corauth[cor1]{Corresponding author.}
\ead{kyo@awa.tohoku.ac.jp}
\author[RCNS]{K. Inoue},
\author[RCNS]{K. Owada},
\author[RCNS]{F. Suekane},
\author[RCNS]{A. Suzuki\thanksref{now}},
\author[TEPCOSYSTEMS]{G. Hirano},
\author[TEPCOSYSTEMS]{S. Kosaka},
\author[TEPCO]{T. Ohta} and
\author[TEPCO]{H. Tanaka}

\address[RCNS]{Research Center for Neutrino Science, Tohoku University, Sendai 980-8578, Japan}
\thanks[now]{Present address: KEK, High Energy Accelerator Research Organization, Tsukuba 305-0801, Japan.}
\address[TEPCOSYSTEMS]{TEPCO Systems Corporation, Tokyo 135-0034, Japan}
\address[TEPCO]{Tokyo Electric Power Company, Tokyo 100-8560, Japan}

\begin{abstract}
KamLAND is a reactor neutrino oscillation experiment with a very long baseline.
This experiment
successfully measured oscillation phenomena of reactor antineutrinos coming
mainly from 53 reactors in Japan.
In order to extract the results, it is necessary to accurately calculate
time-dependent antineutrino spectra from all the reactors.
A simple model of reactor cores and code implementing it were developed for 
this purpose.
This paper describes the model of the reactor cores used in the KamLAND reactor
analysis.
\end{abstract}

\end{frontmatter}

\section{Introduction}

The KamLAND ({\it Kam}ioka {\it L}iquid Scintillator {\it A}nti-{\it N}eutrino
{\it D}etector) experiment \cite{ref:kamland} is 
a reactor neutrino oscillation experiment with a very long baseline.
The experiment detects low energy antineutrinos 
($\overline{\nu}_{\mathrm{e}}$'s) coming from a number of reactors
at a typical distance of $\sim$180 km away from the detector.
KamLAND showed evidence for reactor 
$\overline{\nu}_{\mathrm{e}}$ disappearance 
in 2003 \cite{ref:kamland_1st_result},
and evidence for the distortion of the $\overline{\nu}_{\mathrm{e}}$ energy 
spectrum, which is consistent with neutrino oscillation, 
in 2005 \cite{ref:kamland_2nd_result}.
In order to extract the oscillation signal,
the KamLAND experiment compares the observed and expected energy spectra.
Due to the effect of neutrino oscillation,
the $\overline{\nu}_{\mathrm{e}}$ energy spectrum changes as follows 
in the two-neutrino mixing case,
\begin{equation}
  n(E_{\overline{\nu}_{\mathrm{e}}})=n_{0}(E_{\overline{\nu}_{\mathrm{e}}})\left( 1 - \mathrm{sin}^{2}2\theta 
\,\mathrm{sin}^{2}\frac{\Delta m^{2}L}{4E_{\overline{\nu}_{\mathrm{e}}}} \right),
\end{equation}
where
$E_{\overline{\nu}_{\mathrm{e}}}$ is the $\overline{\nu}_{\mathrm{e}}$ energy,
$n_{0}(E_{\overline{\nu}_{\mathrm{e}}})$ is the $\overline{\nu}_{\mathrm{e}}$
energy spectrum in the absence of neutrino oscillation and
$n(E_{\overline{\nu}_{\mathrm{e}}})$ is the expected 
$\overline{\nu}_{\mathrm{e}}$ energy spectrum with neutrino oscillation.
$\Delta m^2$ is the difference of the squared neutrino masses,
$\Delta m^2 = m_2^2-m_1^2$, 
and $L$ is the distance between the detector and the source.
The oscillation parameters, $\Delta m^2$ and $\sin^22\theta$, are determined 
from the frequency and amplitude of this change. 
The time dependence of the number of $\overline{\nu}_{\mathrm{e}}$'s 
per unit power generation changes as much as 10\%
because the components of nuclear fuel change during the burnup. 
It is therefore necessary to trace the burnup effect for each reactor. 
However, for the case of KamLAND, there are many reactors that 
contribute $\overline{\nu}_{\mathrm{e}}$ events and
it is practically impossible to calculate the burnup effect 
using a detailed simulation for all the reactors.
This paper describes a simple reactor model with which to accurately calculate
the $\overline{\nu}_{\mathrm{e}}$ spectrum of each reactor 
using the routinely recorded reactor operation parameters. 
The parameters include the time-dependent thermal output, burnup, and 
$^{235}\mathrm{U}$ enrichment of exchanged fuel and its volume ratio.    

\section{Antineutrino generation in reactors}

\subsection{Expected reactor antineutrino spectrum}

Nuclear reactors are rich $\overline{\nu}_{\mathrm{e}}$ sources.
The fission products are generally neutron rich nuclei and undergo
successive $\beta$-decays, each yielding one $\overline{\nu}_{\mathrm{e}}$.
About 6 $\overline{\nu}_{\mathrm{e}}$'s are produced per fission along with
an average energy release of $\sim 200$ MeV.
A typical reactor operating at 3 GW thermal output produces 
$\sim 7\times 10^{20}\ \overline{\nu}_{\mathrm{e}}$'s per second.
At the same time, nuclear transmutation through neutron absorption
and beta decays of the fuel elements
changes the composition of the nuclear fuel.
In particular, plutonium is created due to neutron capture by 
$^{238}\mathrm{U}$. 
The plutonium production scheme is shown below.
The produced $^{239}$Pu and $^{240}$Pu are also fissile isotopes and 
contribute to the power and $\overline{\nu}_{\mathrm{e}}$ generation;
\begin{equation*}
  ^{238}\mathrm{U}(\mathrm{n},\gamma)^{239}\mathrm{U}
  \xrightarrow[E_{\mathrm{max}}=1.265\mathrm{MeV}]{T_{1/2}=23.5\mathrm{min}}
  {}^{239}\mathrm{Np}
  \xrightarrow[E_{\mathrm{max}}=0.722\mathrm{MeV}]{T_{1/2}=2.357\mathrm{d}}
  {}^{239}\mathrm{Pu}(\mathrm{n},\gamma)^{240}\mathrm{Pu}(\mathrm{n},\gamma)
  {}^{241}\mathrm{Pu}.
\end{equation*}
The four main isotopes, 
$^{235}\mathrm{U}$, $^{238}\mathrm{U}$, $^{239}\mathrm{Pu}$ and
$^{241}\mathrm{Pu}$, contribute more than 99.9\% of the total power generation.
Contributions from other elements can be safely ignored.
The component of nuclear fuel changes as a function of ``burnup'',
which is defined as the time integrated thermal output $W(t)$
per initial nuclear fuel mass $M$,
\begin{equation}
  b = \int_{0}^{t} \frac{W(t)}{M} \mathrm{d}t.
\end{equation}
The commonly used unit is giga-watt$\times$day/ton.
The burnup is a basic parameter which indicates the condition of nuclear fuel.
For constant thermal output operation, 
the burnup is proportional to the operation time.
Typical nuclear fuel resides in the reactor core 
for 3-5 years and total thermal output is designed to be 30 MW/t.
Therefore, at the end of about one year of each operation cycle, 
incremental burnup from the beginning of cycle
(BOC) reaches approximately 10 GWd/t.

The production rate of reactor $\overline{\nu}_{\mathrm{e}}$'s  
is obtained from the fission rates and the $\overline{\nu}_{\mathrm{e}}$ 
energy spectra per fission.
In the KamLAND data analysis, we compare the experimental data 
with the expected neutrino flux of a superposition of all the reactor cores 
in the integral live time period of the detector.
The energy spectrum of the expected reactor
$\overline{\nu}_{\mathrm{e}}$ flux is written as, 
\begin{align}
  \Psi(E_{\overline{\nu}_{\mathrm{e}}}) =
   \sum_{\mathrm{reactor}} \left(1 - P(\Delta m^{2}, \mathrm{sin}^{2}2\theta, E_{\overline{\nu}_{\mathrm{e}}}, L_{\mathrm{reactor}}) \right)
  \frac{1}{4\pi L_{\mathrm{reactor}}^{2}} \notag  \\
   \times \sum_{\mathrm{isotope}} \psi_{\mathrm{isotope}}(E_{\overline{\nu}_{\mathrm{e}}}) 
  \int_{\mathrm{livetime}}\mathrm{d}t f_{\mathrm{reactor}}^{\mathrm{isotope}}(t),
\end{align}
where
$\psi_{\mathrm{isotope}}(E_{\overline{\nu}_{\mathrm{e}}})$ is the 
$\overline{\nu}_{\mathrm{e}}$ energy
spectrum per fission of each fissile isotope,
$L_{\mathrm{reactor}}$ is the distance from the reactor to KamLAND, 
$P(\Delta m^{2}, \mathrm{sin}^{2}2\theta, E_{\overline{\nu}_{\mathrm{e}}}, L_{\mathrm{reactor}})$ is the oscillation
probability, and
$f_{\mathrm{reactor}}^{\mathrm{isotope}}(t)$ is the time-dependent fission 
rate of each isotope in each reactor.

\subsection{Antineutrino detection}

\begin{figure}[htb]
\begin{center}
\includegraphics[height=7cm]{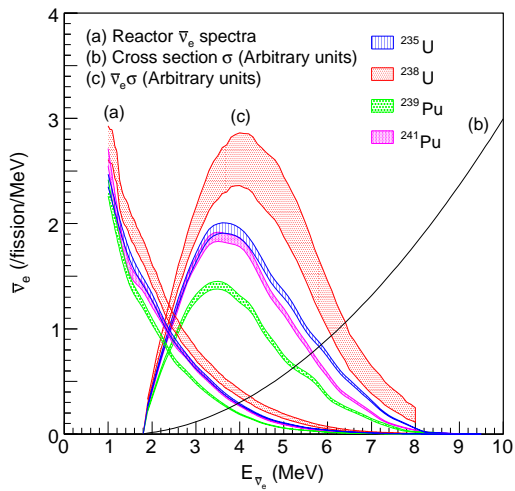}
\caption{(a) Reactor $\overline{\nu}_{\mathrm{e}}$ energy spectra for four main fissile isotopes \cite{ref:neutrino_spectrum}. The shaded region for the isotopes gives the uncertainty in the spectrum. (b) Cross section of the inverse $\beta$-decay reaction \cite{ref:cross_section}. (c) $\overline{\nu}_{\mathrm{e}}$ observed no-oscillation spectrum for each fissile isotope; this is a convolution of (a) and (b).}
\label{fig:test42_each_isotope_graph}
\end{center}
\end{figure}

The KamLAND detector consists of 1 kton of liquid scintillator
surrounded by 1879 17/20-inch-diameter photomultiplier tubes.
In the liquid scintillator, $\overline{\nu}_{\mathrm{e}}$'s are detected 
through the inverse $\beta$-decay reaction with protons,
\begin{equation}
\overline{\nu}_{\mathrm{e}} + \mathrm{p} \rightarrow \mathrm{e}^{+} + \mathrm{n}.
\end{equation}
The cross section is closely related to the neutron lifetime and known with
0.2\% accuracy \cite{ref:cross_section}.
The energy threshold of this reaction is 1.8 MeV and 
the $\overline{\nu}_{\mathrm{e}}$'s produced in the $\beta$-decays of 
$^{239}\mathrm{U}$ and $^{239}\mathrm{Np}$ 
in nuclear fuel do not contribute to the reaction.
The positron and its annihilation gammas produce scintillation light 
which is proportional to $E_{\overline{\nu}_{\mathrm{e}}}-0.8\ \mathrm{MeV}$.
The $\overline{\nu}_{\mathrm{e}}$ visible energy spectrum
can be written as,
\begin{equation}
  n(E_{\overline{\nu}_{\mathrm{e}}}) 
  = N_{\mathrm{p}}\sigma(E_{\overline{\nu}_{\mathrm{e}}})\Psi(E_{\overline{\nu}_{\mathrm{e}}}), \\
\end{equation}
where
$N_{\mathrm{p}}$ is the number of target protons in the detector and
$\sigma(E_{\overline{\nu}_{\mathrm{e}}})$ is the inverse $\beta$-decay 
cross section.
Fig. \ref{fig:test42_each_isotope_graph} shows the neutrino energy spectrum
of each fissile isotope, the inverse $\beta$-decay cross section, 
and the produced visible energy spectra in the detector. 

\begin{table*}[htb]
\begin{center}
\begin{tabular}{|c|cccc|} \hline\hline
                            & \multicolumn{4}{c|}{Isotopes} \\ \cline{2-5}
                            & $^{235}\mathrm{U}$ & $^{238}\mathrm{U}$ & $^{239}\mathrm{Pu}$ & $^{241}\mathrm{Pu}$ \\ \hline
$N_{\overline{\nu}_{\mathrm{e}}}/N_{\overline{\nu}_{\mathrm{e}}}^{{}^{235}\mathrm{U}}$
                           & 1    & 1.52 & 0.60 & 0.87 \\
$\alpha$                   & 0.56 & 0.08 & 0.30 & 0.06 \\
$\xi$                  & 0.61 & 0.13 & 0.20 & 0.06 \\
\hline\hline
\end{tabular}
\caption{Comparisons of the numbers of observed no-oscillation $\overline{\nu}_{\mathrm{e}}$'s per fission above the 3.4 MeV $\overline{\nu}_{\mathrm{e}}$ energy threshold for the isotopes to $^{235}\mathrm{U}$, $N_{\overline{\nu}_{\mathrm{e}}}/N_{\overline{\nu}_{\mathrm{e}}}^{{}^{235}\mathrm{U}}$. $\alpha$ are the contributions of the isotopes to the total number of fission rates in a typical reactor core, corresponding to the contributions to the total energy release. $\xi$ are the contributions of the isotopes to the total number of observed no-oscillation $\overline{\nu}_{\mathrm{e}}$'s above the 3.4 MeV energy threshold from a typical reactor core given from $N_{\overline{\nu}_{\mathrm{e}}}/N_{\overline{\nu}_{\mathrm{e}}}^{{}^{235}\mathrm{U}}$ and $\alpha$.}
\label{table:fission_ratio}
\end{center}
\end{table*}

Above the 3.4 MeV $\overline{\nu}_{\mathrm{e}}$ energy KamLAND analysis 
threshold,
the comparisons of the observed no-oscillation $\overline{\nu}_{\mathrm{e}}$'s
are listed in Table \ref{table:fission_ratio}.
There are significant differences in the number of detected 
$\overline{\nu}_{\mathrm{e}}$'s per fission.
This means that the $\overline{\nu}_{\mathrm{e}}$ spectrum depends on 
the fuel component and burnup. 
The number of observed events from a typical 
reactor core  
decreases by $\sim10\%$ from the BOC to $b=10\mathrm{GWd/t}$. 
Thus when calculating the $\overline{\nu}_{\mathrm{e}}$ flux, 
it is necessary to know the dependence of the fuel components on the burnup.  
The contributions of the isotopes to the total number of fission rates
in a typical reactor core are also listed in Table \ref{table:fission_ratio}.
It must be noted that 
although $^{238}\mathrm{U}$ generates a larger number of neutrinos per fission,
the contribution to the number of observed 
$\overline{\nu}_{\mathrm{e}}$ events is only $\sim 10 \%$.
The shaded regions of the neutrino spectra shown 
in Fig. \ref{fig:test42_each_isotope_graph} indicate the uncertainty 
in the spectra.
Accordingly, the uncertainty in the total number of observed 
$\overline{\nu}_{\mathrm{e}}$'s above the 3.4 MeV energy threshold is 2.5\%.

\subsection{Commercial reactors in Japan}
During the measurement period of KamLAND in Ref. \cite{ref:kamland_2nd_result}
(from 9 March 2002 to 11 January 2004), 
52 commercial reactors 
in 16 electric power stations and a prototype reactor operated in Japan.
All Japanese commercial reactors are light water reactors (LWRs),
29 are boiling water reactors (BWRs) 
and 23 are pressurized water reactors (PWRs).
Both types of LWRs use 3-5\% enriched uranium fuel.
Generally, reactor operation stops once a year 
for refueling and regular maintenance.
During the refueling, one fourth of the total nuclear fuel is exchanged
in BWRs and one third in PWRs.

To calculate production rates of reactor $\bar{\nu}_{e}$'s, 
knowledge of the correlation
between the ``core thermal output'' and the fission rates is 
required.
The core thermal output is defined as the thermal energy generated 
in the reactor cores,
and it is calculated by measuring the heat balance of the reactor cores.
The heat taken out by the cooling water, $Q_{\mathrm{fw}}$, is the dominant 
dissipation source of the reactor energy. 
Other contributions are less than 1\%.
Therefore, the uncertainty of the calculated core thermal output is dominated
by the accuracy of measuring $Q_{\mathrm{fw}}$
which itself
is dominated by the accuracy of measuring the flow of the coolant.
The accuracy of the flow of the coolant in turn is determined by the 
uncertainty of the feedwater flowmeters, which are calibrated to within 2\%.
In the KamLAND experiment, a value of 2\% is used as the uncertainty
of the core thermal output.

All Japanese reactors have a contribution of more than $\sim 0.1$\%
to the total reactor $\overline{\nu}_{\mathrm{e}}$ flux at the location 
of KamLAND
and about half of reactors contribute between 1\% and 7\%.
Therefore, to accurately calculate the total $\overline{\nu}_{\mathrm{e}}$ 
flux in the KamLAND experiment, it is required to trace the time variation of 
the fission rate of all the reactors.
To calculate $f_{\mathrm{reactor}}^{\mathrm{isotope}}(t)$, 
it is necessary to understand the burnup process of nuclear fuel.
The process of burnup is complicated and depends on the type of core,
history of the burnup, initial enrichment, fuel exchange history, etc.
Detailed simulations exist that calculate the change
of the fuel components in accordance with the burnup.
The simulation uses the ``reactor core analysis method'', which
traces the burnup effect of the three-dimensional fuel component 
in the reactor core.
Ideally, it is desirable to perform the simulation for all the reactor cores.
However, it is practically impossible to perform such a detailed simulation
for all the commercial reactors, because it is very labor intensive.
Based on the above circumstances, we have developed a simple
core modeling scheme and calculation code which can be used to calculate
$f_{\mathrm{reactor}}^{\mathrm{isotope}}(t)$ easily without reducing accuracy.
We required that
the discrepancy of the $\overline{\nu}_{\mathrm{e}}$ energy spectrum 
from our simplified model be less than 1\% from the detailed method.
In our model, the burnup effect of the fission rate 
of each fissile isotope in the entire core
is approximated phenomenologically
using reactor operation parameters of the nuclear reactor
and calculated based on a reference reactor core.
All of these parameters have been recorded regularly by the 
nuclear power station, so 
we can use them to calculate the fission rates
without requiring additional effort from the electric power company.
The details of this method are described in the next section.

\section{Fission rate calculation}

\subsection{Reactor core analysis}

\begin{table*}[htb]
\begin{center}
\begin{tabular}{|c|c|l|rr|}\hline\hline
\multicolumn{1}{|c|}{Core number} & \multicolumn{1}{c|}{Core type} & \multicolumn{1}{c|}{Cycle number} & \multicolumn{1}{c}{$\epsilon (\%)$} & \multicolumn{1}{c|}{$V (\%)$} \\ \hline
 1 & BWR  & 1           & 2.2 & 100 \\ \cline{3-5} 
   &      & 2           & 3.0 &  40 \\ \cline{3-5} 
   &      & After 3     & 3.0 &  25 \\ \hline 
 2 & BWR  & 1           & 2.2 & 100 \\ \cline{3-5} 
   &      & 2           & 3.4 &  25 \\ \cline{3-5} 
   &      & After 3     & 3.4 &  29 \\ \hline
 3 & BWR  & 1           & 2.5 & 100 \\ \cline{3-5} 
   &      & 2           & 3.4 &  28 \\ \cline{3-5} 
   &      & After 3     & 3.4 &  26 \\
   &      &             & 3.4 &  28 \\
   &      &             & 3.4 &  29 \\
   &      &             & 3.7 &  26 \\ \hline
 4 & BWR  & After 3     & 3.7 &  26 \\ \hline
 5 & BWR  & After 3     & 3.4 &  23 \\ 
   &      &             & 3.4 &  28 \\ \hline
 6 & PWR  & 1           & 2.2 & 100 \\ \cline{3-5} 
   &      & 2           & 3.5 &  33 \\ \cline{3-5} 
   &      & After 3     & 3.5 &  33 \\ \hline\hline
\end{tabular}
\caption{Parameters of new nuclear fuel in six reference reactor cores used in the present work. Shown are fuel cycle, average enrichment of new fuel $\epsilon (\%)$ and volume ratio of new fuel $V (\%)$.}
\label{table:parameters_of_nuclear_fuel}
\end{center}
\end{table*}

To study the burnup effect of the fission rate in nuclear fuel,
sample reactor cores under actual operating conditions listed in 
Table \ref{table:parameters_of_nuclear_fuel} were analyzed
with the detailed core simulation.
These reactor cores were selected to represent
typical reactor cores in Japan and
the target nuclear fuel was chosen to represent
typical uranium based nuclear fuel
including the initial reactor operation periods.
This analysis used
the Core Management System (CMS) codes from Studsvik of America,
CASMO\cite{ref:CASMO-4}/SIMULATE\cite{ref:SIMULATE-3}.
This system performs core calculations by combining
``two-dimensional multi-group fuel-assembly'' analysis and 
``three-dimensional few-group full core'' analysis.
These codes have been extensively compared with measurements.
The comparison \cite{ref:PHYSOR2002}
of the calculated isotopic concentrations 
provided by the CMS codes with experiments was carried out
for the spent nuclear fuel discharged from 
a BWR type reactor core in Japan.
The fuel pin averaged discrepancies between calculated
and measured isotopic concentrations
are less than 7\% for the four main fissile isotopes.
According to the contributions of the fissile isotopes to the number of
observed $\overline{\nu}_{\mathrm{e}}$'s 
from a typical reactor core listed in Table \ref{table:fission_ratio},
these discrepancies of the isotopic concentrations 
correspond to less than $\sim 1\%$ of
the number of observed $\overline{\nu}_{\mathrm{e}}$'s per 
unit energy release.

\begin{figure}[htb]
\begin{center}
\includegraphics[height=7.0cm]{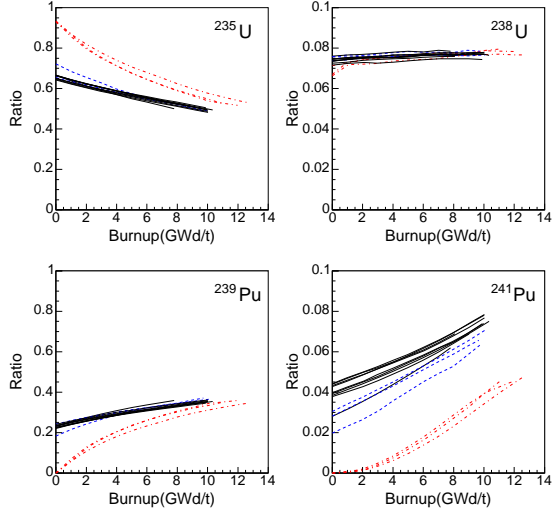}
\caption{Relative fission yields for typical reactor cores using a detailed core composition calculation; The ratio is the fission rate divided by the total fission rate per unit energy generated. Each line is the burnup dependence of a reactor core, multiple cores are shown; the dot-dashed red lines show reactor cores in their initial reactor cycle (after commissioning), the dashed blue lines show reactor cores after the first refueling (in their second cycle) and the solid black lines show cores from the third cycle onwards (equilibrium cores).}
\label{fig:detailed_fission_graph01}
\end{center}
\end{figure}

The results of calculations of the burnup effect on the relative fission rate 
of each fissile isotope in the BWR type reactors
listed in Table \ref{table:parameters_of_nuclear_fuel}
are shown in Fig. \ref{fig:detailed_fission_graph01}.
Hereafter, burnup will be defined as the incremental burnup from the BOC.
From Fig. \ref{fig:detailed_fission_graph01},
it is clear that the dependence of burnup can be classified in two groups:
reactors in their initial fuel cycle, right after commissioning, and 
reactors in their second and higher fuel cycle.
Reactor cores in the initial operation period, which starts without plutonium,
are called ``Initial cores''.
In the initial core, the fission rates of $^{239}\mathrm{Pu}$ increase
rapidly from zero after the BOC,
and fission rates of $^{241}\mathrm{Pu}$ increase gradually following 
$^{239}\mathrm{Pu}$.
After a few fuel exchanges,
the burnup effect in a particular operation period is independent of 
the fuel cycle.
This is called ``Equilibrium core''. 
Because the burnup effect is large in initial cores and small in equilibrium 
cores, the dependence of the burnup effect
on the nuclear fuel parameters is 
calculated separately for the two cases.

\subsection{Initial cores}

\begin{figure}[htb]
\begin{center}
\includegraphics[height=7.0cm]{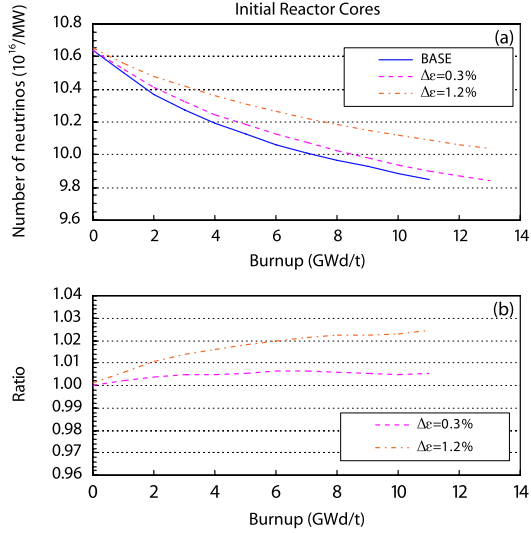}
\caption{(a) Number of reactor $\overline{\nu}_{\mathrm{e}}$'s in typical initial reactor cores calculated with the detailed core simulation. The solid blue line is a core with an enrichment of 2.2\%, the dashed magenta line is 2.5\% and the dot-dashed orange line is a high enrichment core with an enrichment of 3.4\%. (b) Ratio of the number of $\overline{\nu}_{\mathrm{e}}$'s in two typical initial reactor cores to the number of $\overline{\nu}_{\mathrm{e}}$'s in a specific core with an enrichment of 2.2\%.}
\label{fig:system-report-fig3-2}
\end{center}
\end{figure}

The burnup dependence of the number of reactor 
$\overline{\nu}_{\mathrm{e}}$'s between
1.0 MeV and 8.5 MeV $\overline{\nu}_{\mathrm{e}}$ energy produced in typical 
initial reactor cores is shown in Fig. \ref{fig:system-report-fig3-2}.
To extract the dependence on the enrichment, 
a core with a high enrichment of 3.4\% is included in this comparison.
As shown in Fig. \ref{fig:system-report-fig3-2}, 
the number of neutrinos decreases with the burnup 
because of variation of the fuel components.
In highly enriched fuel,
the speed of the decrease is slow. 
It can be explained by the fact that $^{235}\mathrm{U}$ in highly enriched 
fuel has a larger contribution to the total fission rate.
Therefore the enrichment dependence of the burnup effect can be 
represented by transforming the burnup effect in a specific nuclear fuel.

\begin{figure}[htb]
\begin{center}
\includegraphics[height=7.0cm]{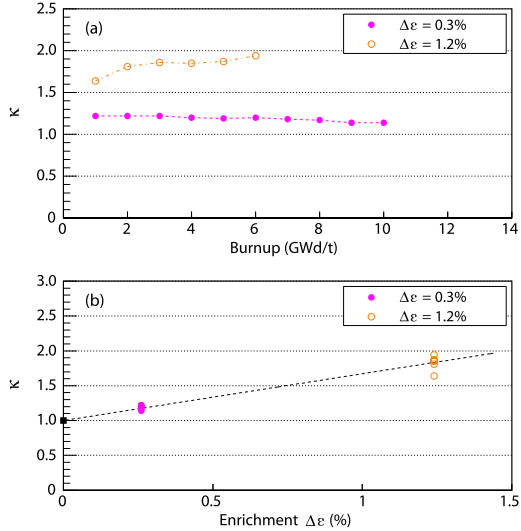}
\caption{(a) Burnup distribution of the $\kappa$ values in two typical initial reactor cores plotted as a perturbation of a specific core with an initial enrichment of 2.2\%; $\kappa$ is approximately constant over the burn cycle. (b) New fuel enrichment dependence of the $\kappa$ values. These values are extracted from the results based on the detailed core calculation method.}
\label{fig:a_graph}
\end{center}
\end{figure}

To study the effect of fuel enrichment, we vary $\epsilon$ to first order,
substituting $\epsilon_{0} \rightarrow \epsilon_{0} + \Delta \epsilon$,
the number of generated $\overline{\nu}_{\mathrm{e}}$'s can be written as
\begin{equation}
N\left(\epsilon_{0}, b\right) =
N\left(\epsilon_{0} + \Delta \epsilon, 
b + \beta(\epsilon_{0}, b) \Delta \epsilon \right),
\end{equation}
where $b$ is the burnup and
\begin{equation}
  \beta(\epsilon_{0}, b) = -
  \frac{\partial N(\epsilon_{0})}{\partial \epsilon} / 
  \frac{\partial N(b)}{\partial b}.
\end{equation}
If we know the value of $\beta(\epsilon_{0}, b)$, the number of 
$\overline{\nu}_{\mathrm{e}}$'s 
can be estimated using a function of burnup for a 
reference value of enrichment, $N(\epsilon_{0}, b)$.
Here, the normalization factor of the burnup, $\kappa$, is defined as,
\begin{equation}
  \kappa = 1 + \frac{\beta(\epsilon_{0}, b)}{b}\Delta \epsilon.
\end{equation}
Fig. \ref{fig:a_graph} shows values of $\kappa$ in typical reactor cores
based on the burnup effect in Fig. \ref{fig:system-report-fig3-2}.
The factor $\kappa$ is approximately constant with respect to the burnup
over the complete fuel cycle (Fig. \ref{fig:a_graph}(a)) and 
can be represented by a linear function of the new 
fuel enrichment (Fig. \ref{fig:a_graph}(b)).
In other words, $\beta(\epsilon_{0}, b)/b$ can be considered to be constant,
and the number of $\overline{\nu}_{\mathrm{e}}$'s is approximated as
\begin{equation}
N\left(\epsilon_{0} + \Delta \epsilon, b\right)
= N\left(\epsilon_{0}, \frac{b}{\kappa}\right).
\end{equation}

In the present work, 
the value of $\beta(\epsilon_{0}, b)/b$ is determined 
by the burnup at the end of cycle ($b_{\mathrm{EOC}}$).
The parameters for the actual reactor conditions
were obtained using the detailed core simulation,
$\kappa = 1 + 0.65 \Delta \epsilon (\%)$ and
$\kappa = 1 + 0.54 \Delta \epsilon (\%)$
for the BWR ($b_{\mathrm{EOC}}=10\mathrm{GWd/t}$) and 
PWR ($b_{\mathrm{EOC}}=12\mathrm{GWd/t}$) cores, respectively.

This simplified reactor core model has so far only been used to calculate
the number of $\overline{\nu}_{\mathrm{e}}$'s.
However, the neutrino oscillation experiment requires
the $\overline{\nu}_{\mathrm{e}}$ energy spectrum.
The burnup effect on the $\overline{\nu}_{\mathrm{e}}$ energy spectrum 
reflects variation of the composition of the fissile isotopes.
We assume for the simple core model that the fission rates of 
all fissile isotopes have the same burnup effect as when calculating
the number of $\overline{\nu}_{\mathrm{e}}$'s.
The model assigns the same corrections for the new fuel enrichment
to the burnup effect to the fission rates.
The accuracy of this treatment is estimated in a later sub-section
by comparing the results from our model to the detailed core simulation.
The systematic uncertainty of the $\overline{\nu}_{\mathrm{e}}$ energy 
spectrum is also described there.

\subsection{Equilibrium cores}

\begin{figure}[htb]
\begin{center}
\includegraphics[height=7.0cm]{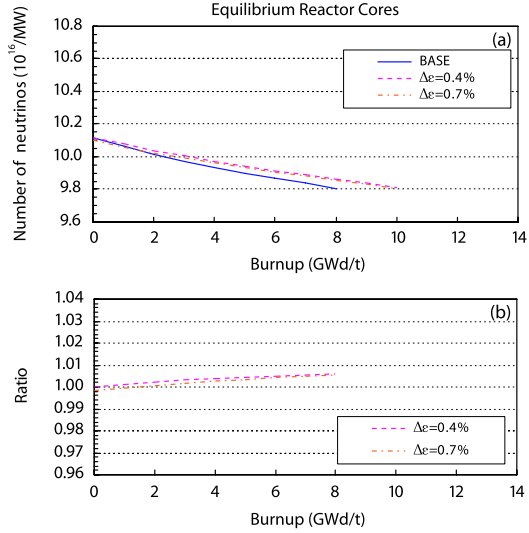}
\caption{(a) Number of reactor $\overline{\nu}_{\mathrm{e}}$'s in typical equilibrium cores calculated with the detailed simulation. The solid blue line is a core with an enrichment of 3.0\%, the dashed magenta line is 3.4\% and the dot-dashed orange line is 3.7\%. (b) Ratio of the number of $\overline{\nu}_{\mathrm{e}}$'s in two typical equilibrium cores to the number of $\overline{\nu}_{\mathrm{e}}$'s in a specific core with an enrichment of 3.0\%.}
\label{fig:Fig3-Equilibrium}
\end{center}
\end{figure}

The burnup dependence of the number of reactor $\overline{\nu}_{\mathrm{e}}$'s
in typical equilibrium cores 
with exchanged fuel volume ratio of approximately 25\% 
is shown in Fig. \ref{fig:Fig3-Equilibrium}.
The relative differences of the number of $\overline{\nu}_{\mathrm{e}}$'s
are already less than 1\% over the complete fuel cycle.
To account for the influence of the new fuel,
we apply the methodology of the fission rate calculation for the 
initial core to the equilibrium core.
Nuclear fuel in equilibrium cores contains both new and old fuel.
The arrangement of new and old nuclear fuel rods is adjusted
to burnup equally in the reactor core.
We treat the fission rate uniformly throughout the core.
Under the assumption that the fission rates of these two components
do not have locational dependence,
the complete fission rate as a function of the burnup of $b$
in an equilibrium core with the exchanged fuel volume ratio $V$
and the new fuel enrichment $\epsilon$
is represented as a combination of the fission rate of the new fuel and of the rest
\begin{align}
  N(b) = &N_{\mathrm{initial}}(b) V + N'(b) \frac{1-V}{1-V_{0}} \\ 
         &N'(b) = N_{\mathrm{equilibrium}}(b) - N_{\mathrm{initial}}(b) V_{0},
  \notag
\end{align}
where $N_{\mathrm{initial}}(b)$ is the estimated fission rate 
in the initial core with the fuel enrichment of $\epsilon$
based on the reference initial core,
and $N_{\mathrm{equilibrium}}(b)$ is the estimated fission rate 
in the equilibrium core with the exchanged fuel volume ratio $V_{0}$ and 
the new fuel enrichment $\epsilon$
based on the reference equilibrium core.
Similarly to the initial core calculation,
$N_{\mathrm{equilibrium}}(b)$ is obtained by 
linear approximation of the burnup effect in a 
reference equilibrium core.
The values of $\kappa$ are
$\kappa = 1 + 0.35 \Delta \epsilon (\%)$ and
$\kappa = 1 + 0.29 \Delta \epsilon (\%)$
for the BWR ($b_{\mathrm{EOC}}=10\mathrm{GWd/t}$) and 
PWR ($b_{\mathrm{EOC}}=12\mathrm{GWd/t}$) cores, respectively.

In addition to the above approximation, 
accumulation of fissile isotopes produced by neutron absorption,
particularly $^{241}\mathrm{Pu}$ produced from the U-Pu chain,
depends on the averaged absolute burnup of nuclear fuel at the BOC.
Correction to the fission rate of each isotope is performed 
as an additional contribution 
using the averaged absolute burnup at the BOC, $b_{\mathrm{absolute}}$
\begin{equation}
  {f'}_{\mathrm{equilibrium}}^{\mathrm{isotope}}(b) = f_{\mathrm{equilibrium}}^{\mathrm{isotope}}(b) + \Delta f^{\mathrm{isotope}}(b_{\mathrm{absolute}}).
\end{equation}
where $f_{\mathrm{equilibrium}}^{\mathrm{isotope}}(b)$ is 
the fission rate of each isotope for the incremental burnup from the BOC, $b$.
The correction term is defined as
\begin{equation}
  \Delta f^{\mathrm{isotope}}(b_{\mathrm{absolute}}) \equiv \eta^{\mathrm{isotope}} \left( \frac{b_{\mathrm{absolute}}}{b^{0}_{\mathrm{absolute}}} - 1 \right),
\end{equation}
where $\eta^{\mathrm{isotope}}$ is the correction factor for each fissile 
isotope and $b^{0}_{\mathrm{absolute}}$ is the averaged absolute burnup 
at the BOC in the reference equilibrium core.
In actual reactor cores, the averaged absolute burnup at the BOC is limited to
$(b_{\mathrm{absolute}}/b_{\mathrm{absolute}}^{0}) \lesssim 1.4$.
The effect of this correction to the fission rate is 
$\lesssim 5\%$ for all fission isotopes except for $^{241}\mathrm{Pu}$
in BWRs, for which the fission rate is corrected by $\lesssim 30\%$.

\subsection{Systematic uncertainties}

\begin{figure}[th]
\begin{center}
\includegraphics[height=7.0cm]{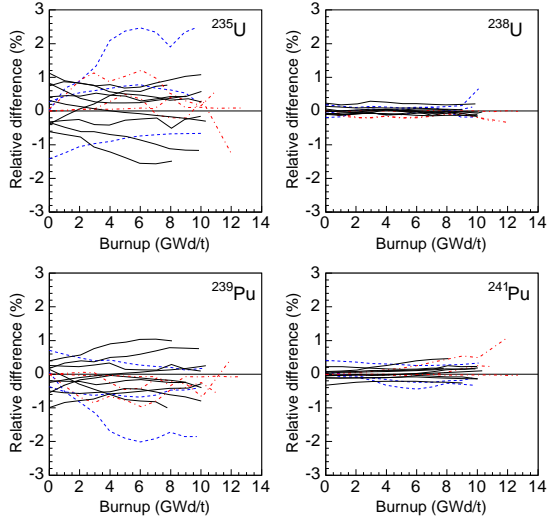}
\caption{Comparison between fission rates in BWR type reactors calculated with the detailed core calculation and our simplified calculation. Each line is the burnup dependence of a reactor core, multiple cores are shown; the dot-dashed red lines show initial cycle reactor cores, the dashed blue lines show second cycle reactor cores and the solid black lines show cores from the third cycle onwards (equilibrium cores). The largest discrepancy between the detailed core calculation and our simplified calculation is for second cycle cores. In practice, this is not of a big concern as most of the 52 commercial reactor cores are already equilibrated and the relative difference is better than 1\%.}
\label{fig:simple_detailed_fission_graph01_for_thesis}
\end{center}
\end{figure}

\begin{figure}[th]
\begin{center}
\includegraphics[width=14.0cm]{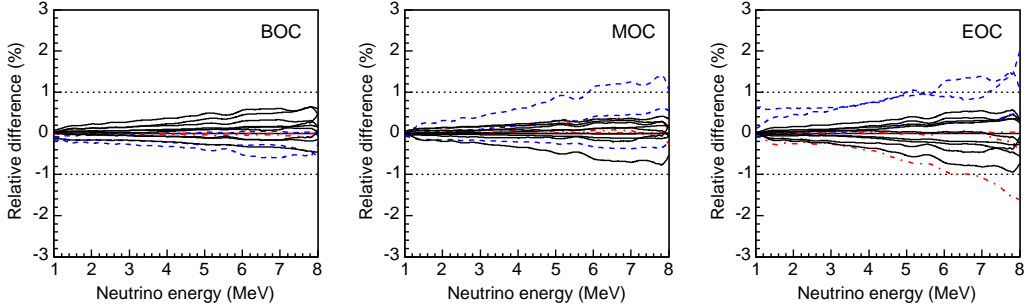}
\caption{Comparison of the $\overline{\nu}_{\mathrm{e}}$ energy spectrum between the detailed core calculation and our simplified calculation in BWR type reactors at the beginning of cycle (BOC), middle of cycle (MOC) corresponding to $\sim 5 \mathrm{GWd/t}$ and end of cycle (EOC). Each line is the burnup dependence of a reactor core, multiple cores are shown; the dot-dashed red lines show initial cycle reactor cores, the dashed blue lines show second cycle reactor cores and the solid black lines show cores from the third cycle onwards (equilibrium cores).}
\label{fig:system-report-fig4-4-2}
\end{center}
\end{figure}

To estimate the accuracy of our simplified calculation method,
we compare our calculations with the detailed reactor core analysis 
performed using parameters from commercial reactors listed in Table 
\ref{table:parameters_of_nuclear_fuel}.
The relative differences in the calculated fission rates 
of each fissile isotopes in BWR type reactors are shown in Fig. 
\ref{fig:simple_detailed_fission_graph01_for_thesis}.
In this figure, the difference is defined as 
$(f_{\mathrm{simplified}}^{\mathrm{isotope}}(b)-f_{\mathrm{detailed}}^{\mathrm{isotope}})(b)/\sum_{\mathrm{isotope}}f_{\mathrm{detailed}}^{\mathrm{isotope}}(b)$.
The difference of the detailed method and simple method is less 
than 3\%.
At the same time, 
the relative differences in the $\overline{\nu}_{\mathrm{e}}$ energy spectra
for different values of burnup in BWR type reactors
are shown in Fig. \ref{fig:system-report-fig4-4-2}.
The $\overline{\nu}_{\mathrm{e}}$ energy spectrum 
in the equilibrium cores agrees with the detailed method within 1\%.
Because the fission rate of each isotope per unit energy generated 
is balanced between the isotopes,
differences in the $\overline{\nu}_{\mathrm{e}}$ energy spectrum
are suppressed.
The discrepancies are large in the high energy region 
in a initial cycle core and second cycle cores.
The maximum discrepancy is 2.0\% in a second cycle core.
However, due to the large number of existing Japanese reactors,
the $\overline{\nu}_{\mathrm{e}}$ contribution from 
second cycle cores is small at the KamLAND experiment and can be neglected.
The reactor $\overline{\nu}_{\mathrm{e}}$ flux calculation is well 
performed with an accuracy of 1\% 
in the number of observed $\overline{\nu}_{\mathrm{e}}$'s 
in all reactor cores, 
which is smaller than the uncertainty of 
the $\overline{\nu}_{\mathrm{e}}$ spectra and the reactor core thermal output.

\section{Summary}
To calculate $\overline{\nu}_{\mathrm{e}}$ flux, reactor cores can be
successfully modeled using only a few reactor operation parameters.
The results of our simplified reactor model agree with
detailed reactor core simulations within 1\% 
for different reactor types and burnup.
This error is taken into account in the KamLAND reactor neutrino analysis.
The simplified model may be applicable to future long-baseline
reactor neutrino experiments which make use of several reactors.

\section{Acknowledgements}
The authors would like to thank M. P. Decowski for helpful discussions 
on this paper.
This work was supported by the Center of Excellence program of the 
Japanese Ministry of Education, Culture, Sports, Science and 
Technology.
The authors gratefully acknowledge the funding support.

\end{document}